\documentclass{ws-procs975x65}

\begin{document}

\title{DOES BLACK HOLE SPIN PLAY A KEY ROLE IN THE FSRQ/BL LAC DICHOTOMY?}

\author{BANIBRATA MUKHOPADHYAY}

\address{Department of Physics, Indian Institute of Science,
Bangalore 560012\\
E-mail: bm@physics.iisc.ernet.in\\
}

\begin{abstract}
It is believed that jets emerging
from blazars (flat spectrum radio quasars
(FSRQs) and BL Lacs) are almost aligned to the line-of-sight. 
BL Lacs usually exhibit lower luminosity
and harder power law spectra at gamma-ray energies than FSRQs.
It was argued previously
that the difference in accretion rates is mainly responsible for the
large observed luminosity mismatch between them. However, when
intrinsic luminosities are derived by correcting for beaming effects,
this mismatch is significantly reduced. 
We show that spin plays an important role to reveal the dichotomy of
luminosity distributions between BL Lacs and FSRQs, suggesting
BL Lacs to be low
luminous and slow rotators compared to FSRQs.

\end{abstract}


\bodymatter

\section{Introduction}\label{intro}

Blazars are the extragalactic radio-loud sources identified in gamma-ray catalogs
of EGRET and Fermi. They are divided by two classes: BL Lacertae (BL Lac) and 
Flat Spectrum Radio Quasar (FSRQ). While the later class shows steep spectra
in high energy and
higher luminosity, the 
former one exhibits flat $\gamma$-ray photon spectral index with a lower
luminosity. Ghisellini et al.\cite{ghis} argued that the luminosity difference 
arises due to difference in accretion rates between the respective systems. 
However, based on MHD simulations, Tchekhovskoy et al.\cite{sasa} 
showed that luminosity dichotomy between radio-loud and radio-quiet Active Galactic
Nuclei (AGNs) arises due to the spin difference between the respective sources.

It has been shown by this author\cite{bgm,bmmg13} that the outflow power and then
luminosity of a black hole system increases with increasing spin of the 
black hole. Based on this theory,
below we establish that the dichotomy between BL Lacs and FSRQs is
due to their spin difference. Moreover, we show that 
the incorporating idea of difference in accretion rates between BL Lacs and FSRQs
gives misleading results.

\section{Basic data and unbeamed luminosities}\label{data}

\begin{figure}[]%
\begin{center}
\vskip-2.2cm 
\hskip-9.6cm 
 \parbox{2.0in}{\epsfig{figure=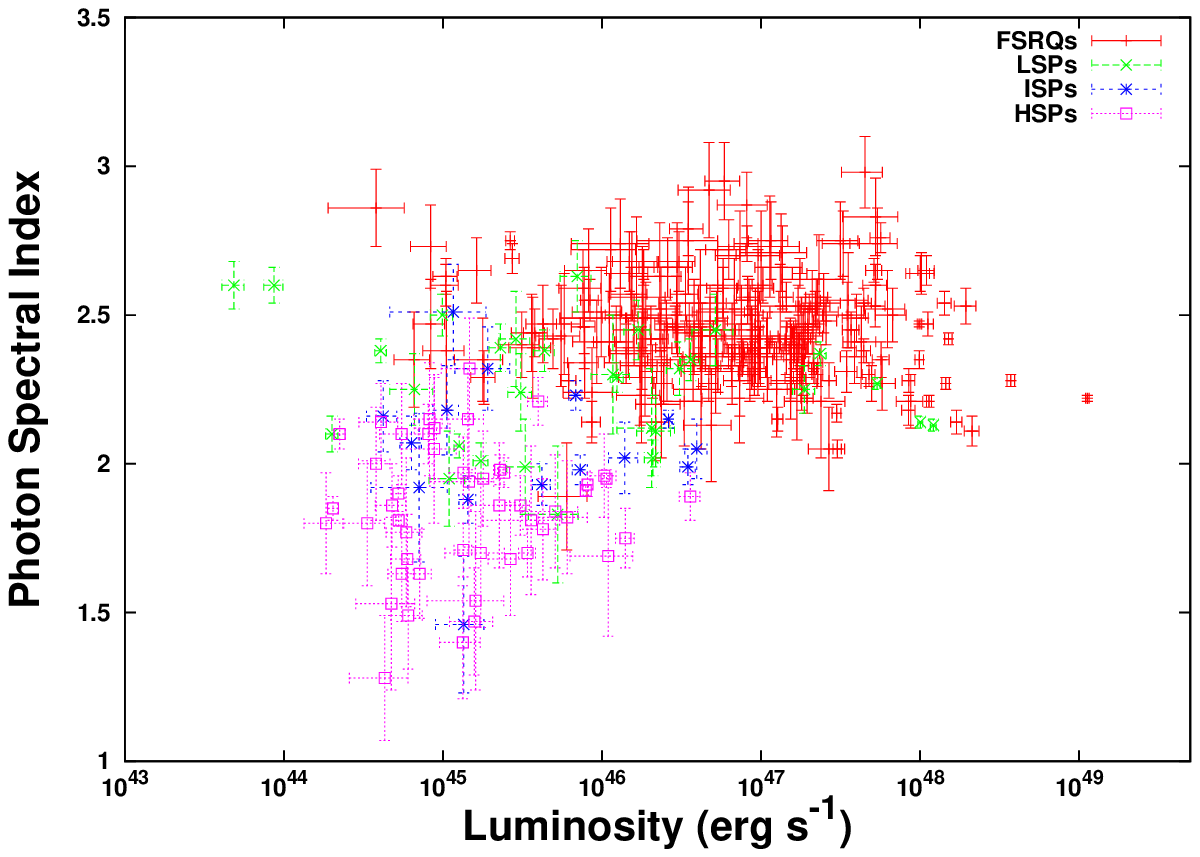,width=3.0in}
%\figsubcap{a}
}
 \hspace*{4pt}
\vskip0.2cm 
\hskip-9.9cm 
 \parbox{2.0in}{\epsfig{figure=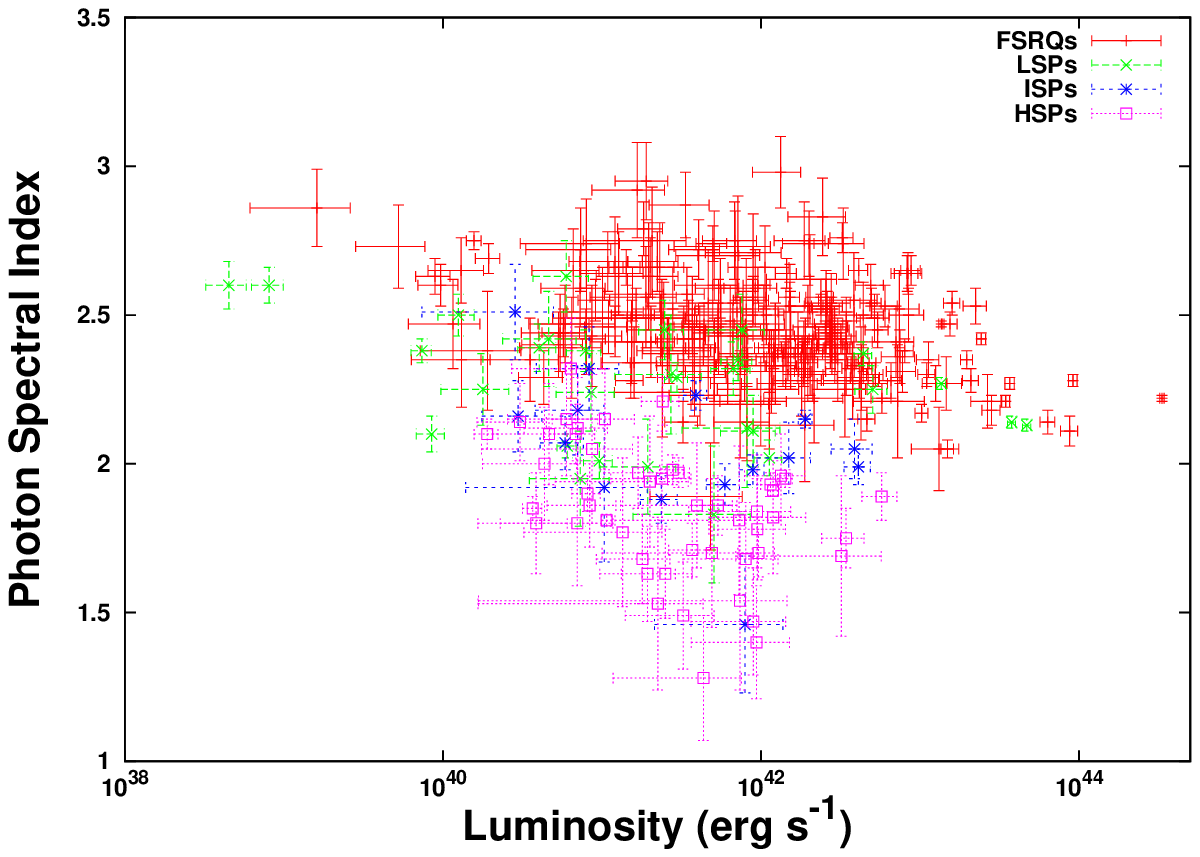,width=3.0in}
% \figsubcap{b}
}
 \caption{\small Upper-Left (a): Observed luminosities and spectral indices for 
FSRQs (red points) and BL Lacs (green, blue and violet points). 
Lower-left (b): Same as in (a), except intrinsic luminosities are plotted;
$50\%$ SSC and $50\%$ EC are chosen for FSRQs.
Upper-right (c): Total mechanical outflow power as a function of black hole's spin,
when solid line represents the fitting function given by equation (\ref{pa}). 
Lower-right (d): The values of FSRQ
spin for a range of BL Lac spin. 
}
%\vskip1cm
\label{fig}
\end{center}
\end{figure}

We consider $11$ months Fermi data\cite{abdo} having $281$ FSRQs and $291$ BL Lacs
with measured redshift ($z$). Further, from radio observations\cite{savo} we 
compute average Doppler beaming factor $\delta\sim 20.6\pm8.4$ of sources. As 
BL Lacs and FSRQs are subclasses of FR~I and FR~II galaxies respectively, when
the jet to line-of-sight angles are very small and hence the jets appear
very beamed, we can relate observed and intrinsic luminosities as\cite{dermer}
\begin{eqnarray}
%\nonumber
L_{\rm observed}=L_{\rm intrinsic}\,\delta^{m+n},
\label{lum}
\end{eqnarray}
where $m=2$ and $3$ for continuous and discrete jets respectively, $n=\alpha_\gamma$ and
$2\alpha_\gamma+1$ for emissions due to synchrotron self-Comptonization (SSC) and 
external Comptonization (EC) processes respectively, $\alpha_\gamma$ is the energy spectral index.
Note that $L_{\rm intrinsic}$ is the quantity produced at the base of jet, 
before it was beamed\cite{bgm}. Hence, this
can be obtained, by knowing $L_{\rm observed}$ and $\delta$ from observed data. 
On the other hand,  $L_{\rm intrinsic}$ can be computed independently from the 
model in Ref.~\refcite{bgm}. Figure \ref{fig}a
shows FSRQs exhibiting on average three orders of magnitude higher 
luminosity and larger photon spectral index than those of BL Lacs. Moreover, FSRQs show
strong emission lines, whereas BL Lacs do not. However, Fig. \ref{fig}b shows
that the difference between respective unbeamed/intrinsic luminosities, from eqn. (\ref{lum}), is
very small. 
Note that $L_{\rm intrinsic}$ is determined solely by the parameters
of disk-outflow coupled dynamics, e.g. accretion rate, mass and spin of the black hole.
From Refs.~\refcite{bgm,bmmg13}, it can be shown that outflow powers and
then $L_{\rm intrinsic}$s between
%will be different between the sources with different 
%accretion rates. Accordingly, if we invoke the earlier suggestion of high difference
FSRQs and BL Lacs\cite{ghis} are significantly different, if their accretion rates are 
also significantly different, as was suggested earlier\cite{ghis},
which contradicts data.
We propose that the difference in spin ($a$) of the underlying 
black holes governs any small difference in $L_{\rm intrinsic}$s between
FSRQs and BL Lacs. 
%FSRQs on average contain higher spinning black holes than BL Lacs. 

%\vskip8cm

\begin{figure}[]%
\begin{center}
\hskip27.1cm 
\vskip-14.6cm
 \parbox{-2.3in}{\epsfig{figure=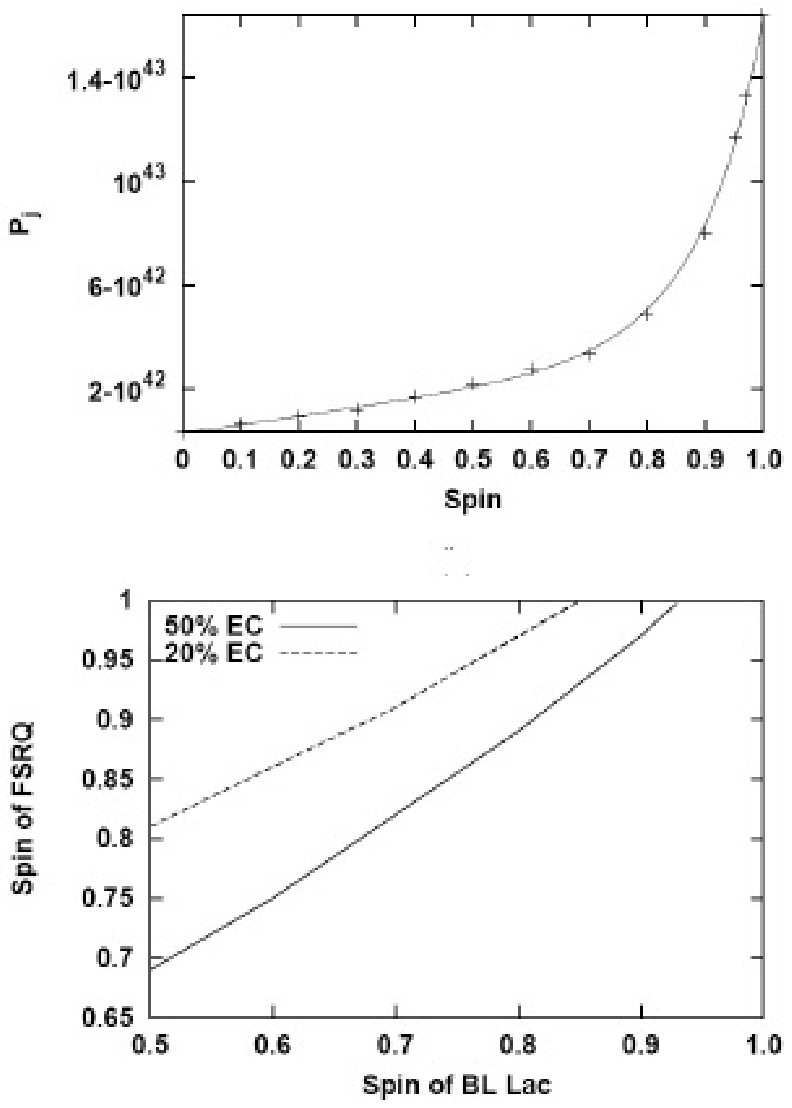,width=3.1in}
% \parbox{-2.3in}{\epsfig{figure=fagn.eps,width=3.2in}
% \figsubcap{c}
}
% \parbox{1.3in}{\epsfig{figure=spin_power.eps,width=1.5in}
% \figsubcap{d}
%}%
% \caption{\small
%}
%\nolabel
%\end{center}
%\end{figure}
%\def\figsubcap#1{\par\noindent\centering\footnotesize(#1)}
%\begin{figure}[b]%
%\begin{center}
%  \parbox{2.1in}{\epsfig{figure=procs-fig2a.eps,width=2in}\figsubcap{a}}
%  \hspace*{4pt}
%  \parbox{2.1in}{\epsfig{figure=procs-fig2b.eps,width=2in}\figsubcap{b}}
%  \caption{Here are two figures side-by-side. (a) Figure caption for figure 2a. (b) Figure caption for figure 2b.}%
%  \label{fig1.2}
\end{center}
\end{figure}

\section{Explaining intrinsic dichotomy via spin difference}\label{dic}

It is found\cite{bgm,bmmg13} that the power of astrophysical jets increases
with increasing spin of central object. We propose a fitting function for jet power
\begin{equation}
P_j =10^{Aa^3+Ba^2+Ca+D},
\label{pa}
\end{equation}
where $A= 2.87{\pm}0.26$, $B= -4.08{\pm}0.40$, $C= 2.88{\pm}0.17$ and $D= 41.53{\pm}0.02$.
As $P_j$ is calculated in the frame of disk-outflow coupled system, we assume in the first approximation
that it is proportional to $L_{\rm intrinsic}$ such that
\begin{eqnarray}
\frac{P_{j,FSRQ}}{P_{j,BLLac}} = \frac{L_{\rm unbeamed,FSRQ}}{L_{\rm unbeamed,BLLac}}, 
\label{ssp}
\end{eqnarray}
but not the observed luminosity.
%due to the inadequate knowledge of the underlying radiation hydrodynamics therein.
Now if we know the spin of BL Lac ($a_{BL}$), then from eqn. (\ref{ssp}) 
the spin of FSRQ ($a_{FSRQ}$) can be obtained, when R.H.S. of eqn. (\ref{ssp}) is known
from observation and denominator and numerator of L.H.S. are respectively supplied 
and computed based on eqn. (\ref{pa}).
Figure \ref{fig}c shows the variation of $P_j$ with $a$.
Now being radio-loud AGNs, blazars are expected to be harboring fast rotating black holes 
than the radio quiet AGNs \cite{sasa} and hence $a\ge 0.5$ in blazars 
(and hence of BL Lacs, which are expected to have relatively
lower luminosity and hence lower $a$ than FSRQs).
Similarly, maximum possible $a$ corresponds to FSRQs. Based on this idea,
Fig. \ref{fig}d 
argues for a range of $a_{FSRQ}$ for a given range of $a_{BL}$ in accordance with the dichotomy in 
luminosities in Fig.  \ref{fig}b.
 \\ \\ \\ \\ \\ \\ \\ \vskip0.7cm

\section{Conclusions}\label{conc}

The observed dichotomy between BL Lacs and FSRQs is due to difference in the underlying spins of 
black holes.
We propose $a_{FSRQ}>a_{BL}$, which is consistent
with their difference in $L_{\rm intrinsic}$s. The small difference in $L_{\rm intrinsic}$s
further gets amplified due to their different emission mechanisms and 
beaming effects, which then leads to large difference in $L_{\rm observed}$s. For further details, 
see Ref.~\refcite{mbs}. \\

%\section*{Acknowledgments}
This work was partially supported by the grant ISRO/RES/2/367/10-11.

%\section{References}
%References are to be listed in the order cited in the text in Arabic
%numerals. \btex users, please use our bibliography style file
%\verb|ws-procs975x65.bst| for references.

%\begin{verbatim}

%\end{verbatim}

%\appendix{About the Appendix}
%Appendices should be used only when absolutely necessary. They
%should come before the References.

%\subappendix{Appendix Sectional Units}

%\bibliographystyle{ws-procs975x65}
%\bibliography{ws-pro-sample}

\begin{thebibliography}{9}
%\bibitem{jarl88} C. Jarlskog, in {\it CP Violation} (World
%                 Scientific, Singapore, 1988).
%\bibitem{lamp94} L. Lamport, {\it \LaTeX, A Document Preparation
%                 System}, 2nd edition (Addison-Wesley, Reading,
%                 Massachusetts, 1994).
%\bibitem{ams04} \AmS-\LaTeX{} Version 2 User's Guide (American
%                Mathematical Society, Providence, 2004).
\bibitem{ghis} G. Ghisellini, L. Maraschi \& F. Tavecchio, {\em MNRAS} {\bf  396}, L105 (2009).
\bibitem{sasa} A. Tchekhovskoy, R. Narayan \& J. C. McKinney, {\em ApJ} {\bf 711}, 50 (2010).
\bibitem{bgm} D. Bhattacharya, S. Ghosh \& B. Mukhopadhyay, {\em ApJ} {\bf 713}, 105 (2010).
\bibitem{bmmg13} B. Mukhopadhyay, {\em this volume}.
\bibitem{abdo} A. A. Abdo, et al., {\em ApJS} {\bf 188}, 405 (2010).
\bibitem{savo} T. Savolainen, et al., {\em A\&A} {\bf 512}, 24 (2010).
\bibitem{dermer} C. D. Dermer, {\em ApJ} {\bf 446}, L63 (1995).
\bibitem{mbs} B. Mukhopadhyay, D. Bhattacharya \& P. Sreekumar {\em IJMPD} {\bf 21}, 1250086 (2012).




%\bibitem{chur90} R.~V. Churchill and J.~W. Brown, {\em Complex
%                 Variables and Applications}, 5th edn.
%                 (McGraw-Hill, 1990).
\end{thebibliography}

\end{document}